\documentclass[12pt,a4paper]{article}
\usepackage{graphicx}
\usepackage{times}
\title{\bf{3-D simulations of shells around massive stars}}
\author{Allard Jan van Marle$^1$ , Rony Keppens$^1$ and Zakaria Meliani$^1$\\
\vspace{1cm}\\
\normalsize $^1$ Centre for Plasma Astrophysics, K.U. Leuven,Celestijnenlaan 200B, \\
                 B-3001 Heverlee, Belgium} 
\date{\mbox{}}
\begin{document}
\maketitle
\pagestyle{empty}
%
% WE REDEFINE THE plain LaTeX PAGESTYLE !!! 
% THIS PAGESTYLE WILL BE USED FOR THE FIRST PAGE ONLY !
%
\def\bull{\vrule height .9ex width .8ex depth -.1ex}
\makeatletter
\def\ps@plain{\let\@mkboth\gobbletwo
\def\@oddhead{}\def\@oddfoot{\hfil\tiny\bull\quad
``The multi-wavelength view of hot, massive stars''; 39$^{\rm th}$ Li\`ege Int.\ Astroph.\ Coll., 12-16 July 2010 \quad\bull}%Result
\def\@evenhead{}\let\@evenfoot\@oddfoot}
\makeatother
%
% AND DEFINE OUR MACROS FOR THE REFERENCE LIST
% I.E \beginrefer \refer and \endrefer
%
\def\beginrefer{\section*{References}%
\begin{quotation}\mbox{}\par}
\def\refer#1\par{{\setlength{\parindent}{-\leftmargin}\indent#1\par}}
\def\endrefer{\end{quotation}}
%
% BEGIN THE ABSTRACT CHAPTER WITH \noindent\small, ENCLOSE IT IN A GROUP
% AND BOLDFACE THE TITLE.
%
{\noindent\small{\bf Abstract:} 
As massive stars evolve, their winds change. 
This causes a series of hydrodynamical interactions in the surrounding medium. 
Whenever a fast wind follows a slow wind phase, the fast wind sweeps up the slow wind in a shell, which can be observed as a circumstellar nebula. 

One of the most striking examples of such an interaction is when a massive star changes from a red supergiant into a Wolf-Rayet star. 
Nebulae resulting from such a transition have been observed around many Wolf-Rayet stars and show detailed, complicated structures owing to local instabilities in the swept-up shells. 

Shells also form in the case of massive binary stars, where the winds of two stars collide with one another. 
Along the collision front gas piles up, forming a shell that rotates along with the orbital motion of the binary stars.
In this case the shell follows the surface along which the ram pressure of the two colliding winds is in balance. 

Using the MPI-AMRVAC hydrodynamics code we have made multi-dimensional simulations of these interactions in order to model the formation and evolution of these circumstellar nebulae and explore whether full 3D simulations are necessary to obtain accurate models of such nebulae. 
}
%
% NOW COMES THE MAIN BODY OF THE ARTICLE
%
\vspace{-0.5cm}
\section{Introduction}
The winds of massive stars create complicated structures in the circumstellar medium. 
The morphology of these structures, known as circumstellar nebulae, is determined by the parameters of the progenitor stars, such as wind velocity, mass loss rate, orbital period and eccentricity as well as the evolutionary sequence of the stars. In this paper we explore two ways in which the wind of a massive star can create a shell.

1) By sweeping up an earlier, slower wind, as has been observed in e.g. NGC~6888 (Treffers \& Chu 1982; Marston 1991), BR~48 (Chu, Weis \& Garnett 1999) and Sh~2-308 (Gruendl et al. 2000). 

2) Through collision with the wind of a massive binary companion. Such shells can not be observed directly, due to their much smaller scale, but their presence can be inferred through variability in the lightcurve as in the case of eta-Carinae (Okazaki et al. 2008) or through dust related IR emission along the tail of the resulting spiral structure as in the case of WR~104 (Monnier et al. 2002)

There are other sources for circumstellar shells, such as outbursts in which mass is ejected in a single, explosive event, rather than through a steady wind, but those fall outside the scope of this paper.

Rather than assuming a symmetry axis, we use the {\tt mpi-amrvac} code (Meliani et al. 2007) to make 3-D models of the wind interactions. 
The single star scenario, in which a fast wind sweeps up its slower predecessor has been explored numerically before by Garc{\'i}a-Segura, Langer \& Mac-Low (1996); van Marle, Langer \& Garc{\'i}a-Segura (2005); Freyer, Hensler \& Yorke (2006); Dwarkadas (2007) and others.
However, such simulations have been limited to 2-D models which, though they provide a great deal of insight, cannot fully describe the fundamentally 3-D structure caused by the instabilities in the shell. 
Because a binary system has no intrinsic symmetry axis it cannot be modelled in 2-D, although pioneering work was done by Stevens, Blondin \& Pollock (1992). 
3-D simulations of the interaction of massive star binary winds have been produced by, among others, Folini \& Walder (2000); Parkin \& Pittard (2008) and Pittard (2009). We add to this by studying the morphology of the shell and how it is influenced by stellar wind and orbital parameters for two different binaries: a luminous blue variable (LBV)+O-star and a hydrogen rich Wolf-Rayet (WNL)+O-star. 

\vspace{-0.5cm}
\section{Numerical method}
We use the {\tt mpi-amrvac} code (Meliani et al. 2007), which solves the conservation equations of hydrodynamics on an adaptive mesh refinement (AMR) grid. 
The use of AMR as well as extensive parallellization allows us to do large scale, 3-D computations. 
In our simulations we ignore the effect of gravity and radiative driving, since the stellar winds are moving at their terminal velocity. 
We include the effect of optically thin radiative cooling, using the solar metallicity cooling table from Mellema \& Lundqvist (2002).  Because hot stars stars radiate strongly in the UV, we assume that the wind material is photo-ionized out to a distance of several parsec as predicted by Eq.~5.14 from Dyson \& Williams 1997 for constant density. 
In an expanding wind the ionization will go to even larger distances (van Marle, Langer \& Garc{\'i}a-Segura 2005; Freyer, Hensler \& Yorke 2006).
Hence, we enforce a minimum temperature of 10\,000~K. In any case, temperature in the wind is largely irrelevant since the kinetic energy dominates the thermal energy by several orders of magnitude.

\subsection{Single star wind interaction}
For the single star simulation we use a spherical grid that covers a radius of 5~pc, with opening angles of $22.5^{\mathrm{o}}$ in both the equatorial and latitudinal plane. 
At the lowest level we use a grid of $400\times64\times64$ grid points and allow a maximum of two additional levels of refinement for a maximum effective resolution of $1600\times256\times256$. Refinement is done dynamically based on local density variations. 

As input parameters we use the stellar evolutionary sequence in which a red supergiant (RSG) becomes a Wolf-Rayet (WR) star. 
This involves a transition from a high mass loss ($10^{-4}$\,M$_\odot$/yr), low velocity (10\,km/s) wind to a lower mass loss rate ($10^{-5}$\,M$_\odot$/yr) with a much higher velocity (2000\,km/s).  The RSG wind parameters are similar to those used in previous 2-D simulations by e.g. Garc{\'i}a-Segura, Langer \& Mac-Low (1996) and van Marle, Langer \& Garc{\'i}a-Segura (2005). 
The WR wind has a somewhat lower mass loss rate than in previous models as it has been observed that WR mass loss rates were over-estimated due to clumping (Vink \& de Koter, 2005). 
We initialize the simulation by filling the grid with the slow RSG wind and starting the fast WR wind at the inner radial boundary. 

\subsection{Binary wind interaction}
We run the binary simulation by calculating the positions of the individual stars as a function of time according to Kepler's law, specifying the direction of motion (counter-clockwise in the XY-plane), the orbital period, stellar masses and eccentricity. 
Round each of these positions we define a sphere, $5\times 10^{12}$\,cm in radius, which is filled with free-streaming wind material. 
For the binary parameters we use two different scenarios: a WNL+O binary, the other a LBV+O binary. 
The parameters of the stellar winds, masses and orbital motion are given in table \ref{tab:bin}, based on values from Vink \& de Koter (2002; 2005). 
We do not include the variability of the LBV wind, but only the steady wind in between outbursts. 
This is possible, because we only look at a relatively short timeframe (1 year), over which the LBV wind can be expected to be steady. 

The binary star simulation is done on a Cartesian grid, spanning $2.5\times10^{14}$~cm in the XY-plane (orbital plane) and $2.5\times10^{13}$~cm along the Z-axis. 
At its lowest resolution the grid measures $240\times240\times20$ points. 
We allow two additional levels of refinement. 
The first is dynamically adjusted based on local density variations, creating an effective resolution of $480\times480\times40$. 
The second is enforced only around the two stars, creating high resolution zones in the regions where the stellar winds are initialized. 

 \begin{table}
      \caption[]{Binary parameters. Primary star: subscript 1, secondary: subscript 2.}
         \label{tab:bin}
\begin{center}
\begin{tabular}{|c|cc|cccc|cc|}
\hline\hline
Simulation & Mass$_1$       & Mass$_2$    & $\dot{M}_1$       & $\dot{M}_2$       & $v_1$ & $v_2$ & Period & eccentricity   \\
           & [M$_\odot$]    & [M$_\odot$] & [M$_\odot /yr]$   & [M$_\odot /yr]$   & [km/s]    & [km/s]    & [yr]   &        \\ 
\hline      
WNL+O      &  50            & 30          & $5\times 10^{-6}$ & $5\times 10^{-7}$ &  1500     & 2000      & 1      & 0       \\
LBV+O      &  50            & 30          & $1\times 10^{-4}$ & $5\times 10^{-7}$ &   200     & 2000      & 1      & 0       \\
\hline
\end{tabular}
\end{center}
 \end{table}

\begin{figure}[h]
\begin{minipage}{8cm}
\centering
\includegraphics[width=8cm]{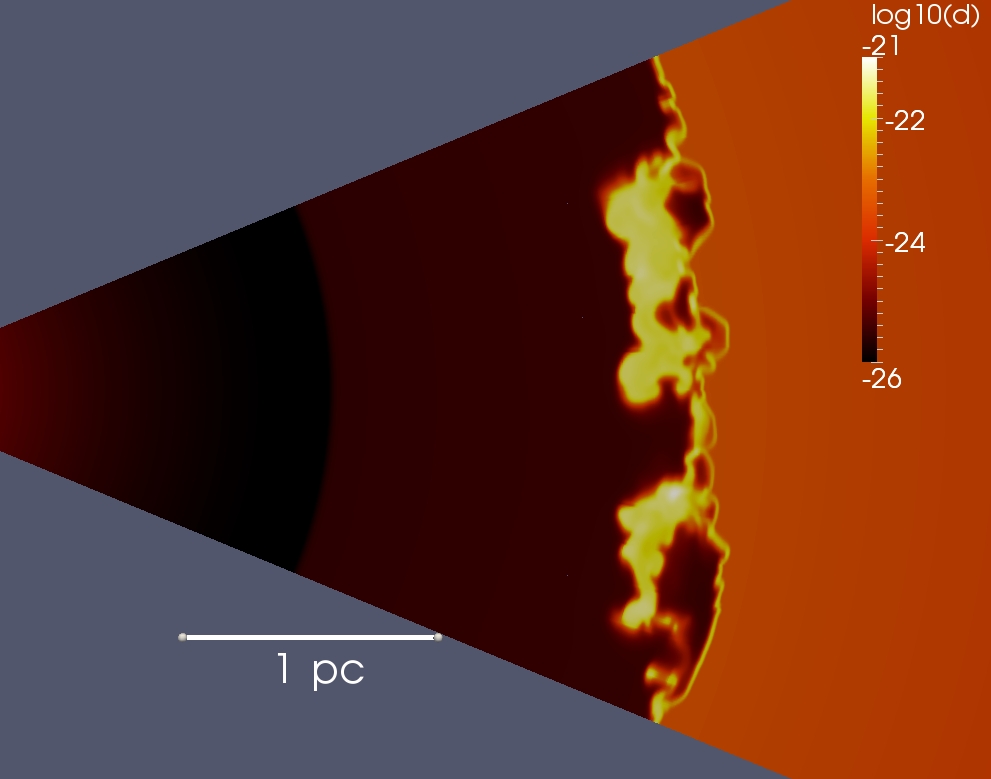}
\caption{Density in g/cm$^3$ for a WR wind sweeping up its RSG predecessor after 40,000 yr. 
This figure shows a 2D slice through the 3D data set. The swept-up shell is thin and show thin-shell instabilities of the linear type.\label{WR_RSG_1}}
\end{minipage}
\hfill
\begin{minipage}{8cm}
\centering
\includegraphics[width=8cm]{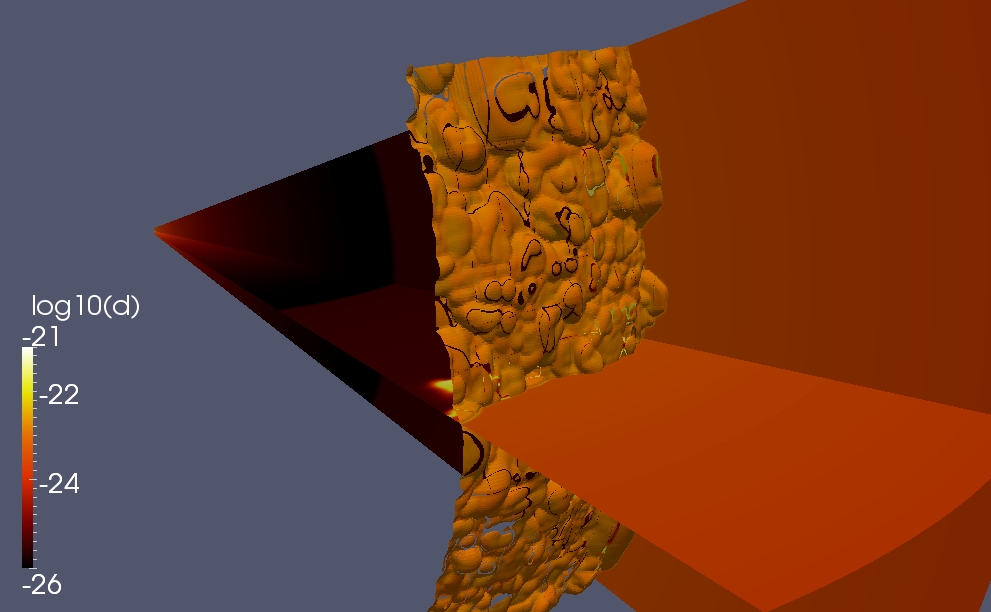}
\caption{Density in g/cm$^3$ for a WR wind sweeping up its RSG predecessor. The shell is shown as an iso-surface at 30\,km/s, which follows the front of the shell. The 3-D structure of the instabilities is clearly visible.\label{WR_RSG_2}}
\end{minipage}
\end{figure}

\vspace{-0.5cm}
\section{Single star circumstellar shell}
The shell of swept-up RSG wind material (Figs.~\ref{WR_RSG_1} and \ref{WR_RSG_2}) after 40,000 yr shows a complicated structure resulting from a combination of thin-shell and Rayleigh-Taylor instabilities as was found in 2-D simulations (Garc{\'i}a-Segura et al. 1996). 
The shell itself is highly compressed because of the radiative nature of the forward shock, which lowers the thermal pressure of the shocked RSG wind material.  Note that the resolution of the grid places a limit on the compression factor, which may influence the results. 
Although the 2-D representation (Fig.~\ref{WR_RSG_1}) strongly resembles earlier results from 2-D simulations, the actual nature of the instabilities is best modeled in 3-D to fully understand their morphology (Fig.~\ref{WR_RSG_2}). 
E.g. The density plot shows a difference in cross-section of about an order of magnitude between the clumps and the shell. Garc{\'i}a-Segura et al. (1996) show only a variation of a factor 2 before the shell breaks out of the RSG wind. 
Since emission of matter depends on the density squared, such structures show up clearly in observations of circumstellar nebulae such as NGC~6888.

\begin{figure}[h]
\begin{minipage}{9cm}
\centering
\includegraphics[width=8cm]{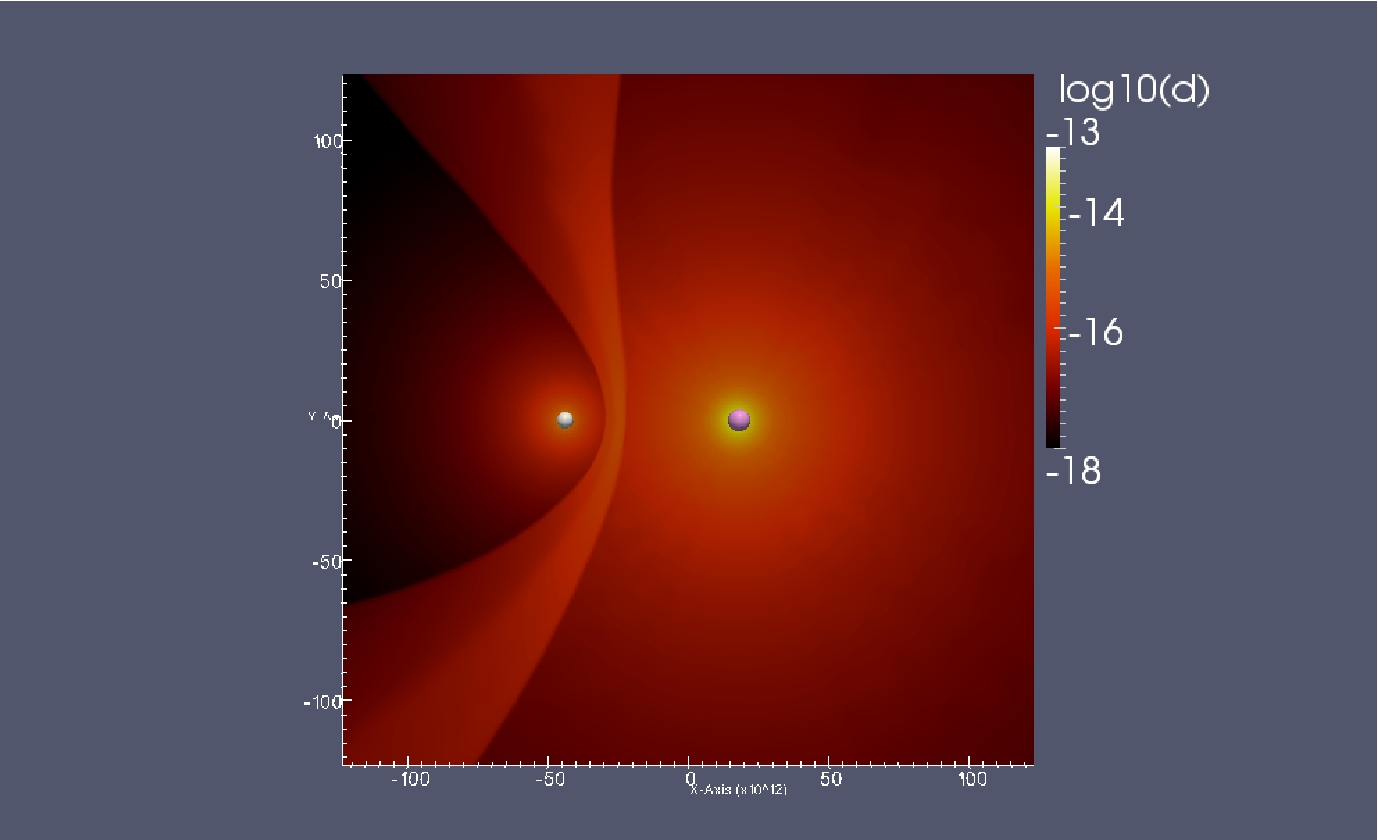}
\caption{Density in g/cm$^3$ in the orbital plane of a WNL+O binary, after one full orbit (1\,yr). The binary stars rotate in counter-clockwise direction. This simulation produces a smooth shell because the wind collisions on both the O-star (white sphere) side and the WNL (violet sphere) side are adiabatic in nature. The resulting shell is thick and not subject to thin-shell instabilities. \label{WNL_O}}
\end{minipage}
\hfill
\begin{minipage}{9cm}
\centering
\includegraphics[width=8cm]{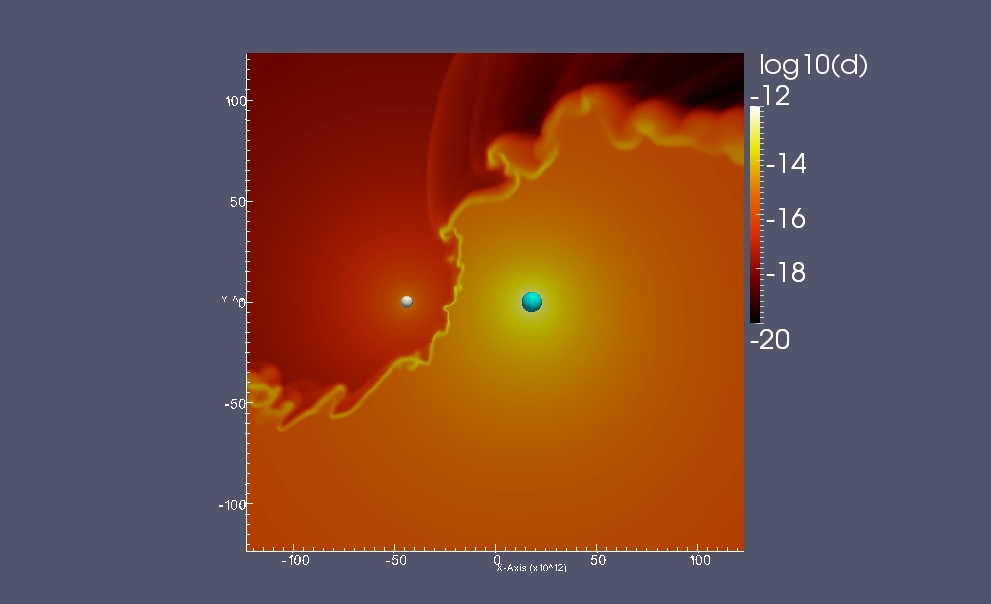}
\caption{Density in g/cm$^3$ in the orbital plane of a LBV+O binary, after one full orbit (1\,yr). Since the wind collision is radiative on the LBV  (blue sphere) side of the collision the shell is thin and subject to instabilities. The trailing end of the shell (top) lags behind due to the low velocity of the LBV wind . \label{LBV_O}}
\end{minipage}
\end{figure}
\vspace{-0.5cm}
\section{Massive binary shells}
The two binary simulations show completely different results (Figs.~\ref{WNL_O} and \ref{LBV_O}), which show slices in the orbital plane through the 3-D dataset. The WNL+O binary has a smooth shell that follows the bowshock curve along which the ram pressure of the two winds is equal, with only a small deviation resulting from the orbital motion. 
The LBV+O binary has a thin, clearly unstable shell. 
The leading edge of the shell, which advances into the LBV wind, follows a general shape similar to that of the WNL+O binary, but the trailing edge (top of Fig.~\ref{LBV_O}) deviates strongly from the bowshock shape, trailing behind the counter-clockwise orbital motion. 
The difference is caused by the wind parameters, which determine whether the wind collision shock is radiative or adiabatic in nature (Stevens, Blondin \& Pollock 1992).  
Owing to the high velocity and (relatively) low density of the WNL and O-star winds, the radiative cooling at the collision is inefficient making the collision nearly adiabatic. The result is a thick shell. 
On the other hand, the LBV wind, with its low velocity and high density creates a radiative shock. 
Therefore, the resulting shell is compressed and susceptible to thin-shell instabilities (Vishniac 1983,1994). 
The strong deviation from symmetry in the global shape of the  LBV+O shell is caused by the relatively low velocity of the LBV wind, relative to the orbital velocity ($v_w$~=~200\,km/s for the LBV wind versus $v_{\rm orbit}~=~40$\,km/s).
Because of its low velocity, the LBV wind cannot push the bowshock ahead fast enough to keep up with the orbital motion of the binary at larger radii. 
As a result the trailing end of the bowshock starts to fall behind leading to the asymmetric shape (Parkin \& Pittard 2008, van Marle, Keppens \& Meliani 2010). 
The WNL wind, which is 7.5 times faster than the LBV wind can push the shell much faster. Consequently, the shell can maintain a near symmetric shape, co-rotating with the orbital motion of the binary stars. 

Comparison with observations is difficult for these systems, because the shells cannot be observed directly. 
Their properties have to be inferred from indirect observation, such as dust trails and variations in the lightcurve. 

\vspace{-0.5cm}
\section{Conclusions}
In order to fully understand the structure end evolution of circumstellar shells, high resolution, 3-D hydrodynamics are necessary. 
For the single star scenario, 2-D models can be used in many applications, but cannot fully capture the details of the unstable shell. 
For the binary scenario 3-D simulations are unavoidable, due to the lack of a symmetry axis. 
Although the colliding winds can be simulated in 2-D using cylindrical symmetry, orbital motion would have to be neglected. 

In both cases radiative cooling plays a crucial part in the shape of the shell as it determines the compression factor of the shock, which in turn determines whether the shell becomes thick and smooth, or thin and subject to instabilities. 
Since we assume photo-ionization, which sets a minimum temperature for the gas, the compression factor is limited. 
Should we allow the gas to recombine and cool further, which may occur if clumps of gas becomes dense enough to shield themselves from the ionizing radiation than the shells would be compressed even further. 

In the future we will explore a larger parameter space of both single star and binary wind interactions in 3-D. 
\vspace{-0.5cm}
\section*{Acknowledgements}
A.J.v.M.\ acknowledges support from FWO, grant G.0277.08 and K.U.Leuven GOA/09/009. 
Simulations were done at the Flemish High Performance Computer Centre, VIC3 at K.U. Leuven and on the CINECA SP6 at Bologna, Italy
We thank the DEISA Consortium (www.deisa.eu), funded through the EU FP7 project RI-222919, for support within the DEISA Extreme Computing Initiative\\
%
% BEGIN THE REFERENCE LIST WITH \beginrefer
% USE \refer BEFORE THE REFERENCES AND BEGIN A NEW PARAGRAPH AFTER THE 
% REFERENCE !
% DO NOT FORGET TO END THE LIST WITH \endrefer
%
\vspace{-0.5cm}
\footnotesize
\beginrefer
\refer Chu, Y.-H., Weis, K, Garnett, D.R.,, AJ, 1999, 117, 1433

\refer Dwarkadas, V.V., 2007, ApJ, 667, 226 

\refer Dyson, J.E. \& Williams, D. 1997, {\it The physics of the interstellar medium}
2$^{nd}$ Edition, (Institute of Physics Publishing)

\refer Folini, D. \& Walder, R., 2000, AP\&SS, 274, 189

\refer Freyer, T., Hensler, F. \& Yorke, H.W., 2006, ApJ, 638, 262

\refer Garc{\'i}a-Segura, G., Langer, N. \& Mac-Low, M.M., 1996, A\&A, 316, 133 

\refer Greundl, R.A., Chu, Y.-H., Dunne, B.C. \& Points, S.D., 2000, AJ, 120, 2670

\refer Marston, A.P., 1991, 366, 181

\refer Meliani, Z., Keppens, R., Casse, F \& Giannios, D., MNRAS, 376, 1189

\refer Mellema, G. \& Lundqvist, P., 2002, A\&A, 394, 901

\refer Monnier, J.D., Greenhill, L.J., Tuthill, P.G. \& Danchi, W.C., 2002, ApJ, 566, 399

\refer Okazaki, A.T., Owocki, S.P., Russel, C.M.P. \& Corcoran, M.F., 2008, MNRAS, 388, L39

\refer Parkin, E.R. \& Pittard, J.M, 2008, MNRAS, 388, 1047

\refer Pittard, J.M., 2009, MNRRAS, 396, 1743

\refer Stevens, I.R., Blondin, J.M. \& Pollock, A.M.T., 1992, ApJ, 386, 265

\refer Treffers, R.R. \& Chu, Y.-H., ApJ, 1982, 254, 569

\refer van Marle, A.J., Keppens, R. \& Meliani, Z. 2010, accepted by A\&A, arXiv:1011.1734

\refer van Marle, A.J., Langer, N. \& Garc{\'i}a-Segura, G., 2005, A\&A, 444, 837

\refer Vink, J.S. \& de Koter, A., 2002, A\&A, 393, 543

\refer Vink, J.S. \& de Koter, A. 2005, A\&A, 442, 587

\refer Vishniac, E.T., 1983, ApJ, 274, 152

\refer Vishniac, E.T., 1994, ApJ, 428, 186

\endrefer
\end{document}